\documentclass[aps,pra,twocolumn,10pt,letterpaper,superscriptaddress]{revtex4-1}

\usepackage{mathptmx,xcolor,graphicx}
\usepackage{amsfonts,amsmath,amssymb}
\usepackage{ulem,enumerate}
\usepackage[colorlinks=true,urlcolor=blue,citecolor=blue,linkcolor=blue]{hyperref}

\begin{document}

\title{Separable and Inseparable Quantum Trajectories}

\author{J. Sperling}\email{jan.sperling@physics.ox.ac.uk}
\affiliation{Clarendon Laboratory, University of Oxford, Parks Road, Oxford OX1 3PU, United Kingdom}

\author{I. A. Walmsley}
\affiliation{Clarendon Laboratory, University of Oxford, Parks Road, Oxford OX1 3PU, United Kingdom}

\date{\today}

\begin{abstract}
	The dynamical behavior of interacting systems plays a fundamental role for determining quantum correlations, such as entanglement.
	In this Letter, we describe temporal quantum effects of the inseparable evolution of composite quantum states by comparing the trajectories to their classically correlated counterparts.
	For this reason, we introduce equations of motions describing the separable propagation of any interacting quantum system, which are derived by requiring separability for all times.
	The resulting Schr\"odinger-type equations allow for comparing the trajectories in a separable configuration with the actual behavior of the system and, thereby, identifying inseparable and time-dependent quantum properties.
	As an example, we study bipartite discrete- and continuous-variable interacting systems.
	The generalization of our developed technique to multipartite scenarios is also provided.
\end{abstract}

\maketitle

\paragraph*{Introduction.---}\hspace*{-3ex}
	The discovery of quantum physics fundamentally altered our understanding of nature \cite{S35}.
	For instance, the phenomenon of quantum entanglement is incompatible with the classical concept of correlations \cite{HHHH09}.
	To classify entanglement, one has to define the notion of separability first.
	For example, a pure bipartite separable state has the form $|a\rangle\otimes|b\rangle=|a,b\rangle$.
	The inseparability of a state defines entanglement, and it was used to object the consequences of quantum physics \cite{EPR35}.
	Another remarkable aspect of quantum physics is that the evolution of particles is described through a wave equation, the Schr\"odinger equation (SE),
	\begin{align}\label{eq:SE}
		i\hbar |\dot\psi\rangle=\hat H|\psi\rangle,
	\end{align}
	where we use $\dot f=df/dt$ for a time-dependent function $f=f(t)$.
	For example, the SE explains the observation of quantum interferences of particles \cite{DG28}, confirming their wave nature.
	In its time-independent form, the SE yields the eigenvalue equation (EE) of the Hamiltonian $\hat H$,
	\begin{align}\label{eq:EE}
		\hat H|\psi\rangle=E|\psi\rangle,
	\end{align}
	which is connected to the quantization of, for example, electronic states in a hydrogen atom \cite{S26}.

	To verify entanglement, a number of inseparability criteria have been formulated \cite{HHHH09,GT09}.
	One of the most successful approaches to certify inseparability are entanglement witnesses \cite{HHH96,HHH01}.
	Therefore, the formulation and optimization of such witnesses have been intensively studied; see, e.g., Refs. \cite{T00,LKCH00,T05,HE06,SRLR17}.
	Among other approaches, the method of so-called separability eigenvalue equations (SEEs) allows for constructing entanglement witnesses \cite{SV09,SV13}.
	The solutions of the SEEs have been used to experimentally identify path-entangled photons \cite{GPMMSVP14} and complex multipartite entanglement in frequency combs \cite{GSVCRTF15,GSVCRTF16}.
	For the purpose of this Letter, it is also worth mentioning that the SEEs have been recently applied to the Hamiltonian to certify entanglement in macroscopic systems \cite{SW17}.
	This progress in verifying inseparability is, however, restricted to a single time or stationary scenarios and does not capture the dynamics of entanglement.

	The interaction between initially separated subsystems typically leads to inseparability; see, e.g., Refs. \cite{FABWR07,LZWLHTY17} for experimental realizations.
	The other way around, the transition of an entangled system to a separable one has been studied as well \cite{YE04,YE09}.
	As a result, a vast number of impressive results have been obtained which describe the evolution of entanglement in various complex systems \cite{WBPS06,KMTKAB08,TMB08,I09,G10,GG12}.
	While these approaches allow us to predict the temporal behavior of inseparability, a comparison of the system with a separable one has to be made for each point in time individually to infer entanglement.
	It is also worth mentioning that witnesses to probe causal inseparability have been recently introduced \cite{ABCFGB15,B16}.

	Therefore, the understanding of the evolution and detection of entanglement has made remarkable progress.
	However, a method to compare the entire evolution (i.e., the trajectory of the system) with a separable one---without excluding interactions---is missing so far.

	In this Letter, we derive equations of motion for compound systems restricted to separable states for arbitrary Hamiltonians with the aim of verifying inseparable trajectories.
	We are able to perform such a task by using the least action principle, which eventually leads to a set of nonlinear and coupled equations for the individual subsystems---to be termed separability Schr\"odinger equations (SSEs).
	We prove fundamental properties of the resulting dynamics and compare it with the actual, i.e., unrestricted, dynamics.
	We solve our equations for an interacting system, which allows us to discern inseparable trajectories from separable ones.
	As a proof of concept, we also demonstrate the generalization to multipartite systems.

\paragraph*{Separability Schr\"odinger equations.---}\hspace*{-3ex}
	To formulate the desired equations of motions, we follow one derivation of the SE \cite{GR96}.
	That is, the least action principle, in which the action $S=\int_0^T dt\,L$ is minimized, is applied using the Lagrangian \cite{comment1}
	\begin{align}
		L=\frac{i\hbar}{2}\langle \psi|\dot\psi\rangle-\frac{i\hbar}{2}\langle\dot\psi|\psi\rangle-\langle \psi|\hat H|\psi\rangle.
	\end{align}
	To include the restriction to separable states, we postulate that we have a product state for all times, $|\psi(t)\rangle=|a(t),b(t)\rangle$.
	This yields Euler-Lagrange equations for the subsystems $A$ and $B$ in the form
	\begin{align}\label{eq:SepEulerLagrange}
		0=\frac{d}{dt}\frac{\partial L}{\partial \langle \dot a|}-\frac{\partial L}{\partial \langle a|}
		\quad\text{and}\quad
		0=\frac{d}{dt}\frac{\partial L}{\partial \langle \dot b|}-\frac{\partial L}{\partial \langle b|}.
	\end{align}
	See Ref. \cite{D12} for an introduction to the calculus of variations.
	Applying the product rule, $\langle\dot\psi|=\langle \dot a,b|+\langle a,\dot b|$, we find
	\begin{align}\label{eq:AuxRel}
	\begin{aligned}
		\frac{\partial L}{\partial \langle \dot a|}
		=&-\frac{i\hbar}{2}\langle b|b\rangle |a\rangle,
		\\
		\frac{\partial L}{\partial \langle a|}
		=&\frac{i\hbar}{2}\big(\langle b|b\rangle|\dot a\rangle+\langle b|\dot b\rangle|a\rangle\big)
		-\frac{i\hbar}{2}\langle \dot b|b\rangle|a\rangle-\hat H_b|a\rangle,
	\end{aligned}
	\end{align}
	using $\hat H_b=\mathrm{tr}_B[\hat H(\hat 1_A\otimes|b\rangle\langle b|)]$ \cite{SV09,SV13} and where $\hat 1_A$ is the identity in $A$ and $\mathrm{tr}_B$ is the partial trace over $B$.
	Inserting relations \eqref{eq:AuxRel} into Eq. \eqref{eq:SepEulerLagrange}, we obtain for $|a\rangle$
	\begin{subequations}
	\begin{align}\label{eq:SSEa}
		i\hbar\big(\langle b|b\rangle |\dot a\rangle+\langle b|\dot b\rangle |a\rangle\big)=\hat H_b|a\rangle
	\end{align}
	and---in analogy---for $|b\rangle$
	\begin{align}\label{eq:SSEb}
		i\hbar\big(\langle a|a\rangle |\dot b\rangle+\langle a|\dot a\rangle |b\rangle\big)=\hat H_a|b\rangle.
	\end{align}
	\end{subequations}

	We refer to this coupled set of Eqs. \eqref{eq:SSEa} and \eqref{eq:SSEb} as SSEs.
	They describe the evolution of a composite system restricted to separable trajectories $|a(t),b(t)\rangle$.
	By construction, this is true for any coupling between $A$ and $B$.
	Comparing the solutions of the SSEs with those of the SE [Eq. \eqref{eq:SE}], we are able to study the inseparable dynamics of compound quantum systems.
	Let us stress that we do not simply take an entangled state $|\psi(t)\rangle$, obtained via a SE, and relate it to separable states.
	Rather, our SSEs describe the separable dynamics itself.

	Let us now characterize the SSEs to understand their physical features.
	The proofs are provided in the Supplemental Material \cite{supplement}.
	These properties are similar to those of the SE and, thus, may not be surprising.
	However, let us emphasize that it is a nonlinear dynamics, described by the SSEs, which exhibits such similarities to the linear propagation of the SE.

	First, the SSEs conserve the normalization; i.e., for all times $t$ holds
	\begin{align}
		\langle a(t)|a(t)\rangle=1
		\quad\text{and}\quad
		\langle b(t)|b(t)\rangle=1.
	\end{align}
	Note that the initial states are assumed to be normalized to one, $\langle a(0)|a(0)\rangle=1=\langle b(0)|b(0)\rangle$.
	Also, the energy of the system subjected to the separable evolution is conserved,
	\begin{align}
		\dot E=0,
	\end{align}
	where $E=\langle a(t),b(t)|\hat H|a(t),b(t)\rangle$.
	Second, we can additionally formulate a von Neumann form of the SSEs,
	\begin{align}\label{eq:vonNeumann}
	\begin{aligned}
		i\hbar\frac{d}{dt}\big(|a\rangle\langle a|\big)&=\big[\hat H_b,|a\rangle\langle a|\big],
		\\
		i\hbar\frac{d}{dt}\big(|b\rangle\langle b|\big)&=\big[\hat H_a,|b\rangle\langle b|\big],
	\end{aligned}
	\end{align}
	which directly compares to the von Neumann equation for the SE, $i\hbar\, d(|\psi\rangle\langle \psi|)/dt=[\hat H,|\psi\rangle\langle \psi|]$.
	Note that this form turns out to be convenient for proving the properties of the SSEs \cite{supplement}.

	Third, in the stationary case, we find that the SSEs are identical to the SEEs of the Hamiltonian,
	\begin{align}\label{eq:SEEs}
		\hat H_b|a\rangle=E|a\rangle
		\quad\text{and}\quad
		\hat H_a|b\rangle=E|b\rangle,
	\end{align}
	see also Refs. \cite{SV09,SV13,supplement}.
	This result is especially interesting in comparison with the SE \eqref{eq:SE}, whose time-independent form is given by the EE \eqref{eq:EE}.
	As mentioned earlier, the SEEs have been used to experimentally detect entanglement \cite{GPMMSVP14,GSVCRTF15,GSVCRTF16} and also for theoretical studies; see Ref. \cite{BSV17} for a recent application.
	For instance, the solutions of Eq. \eqref{eq:SEEs} allowed us to compare the separable spectrum with the actual spectrum of energies of $\hat H$ \cite{SW17}.
	In addition, replacing $\hat H$ in Eq. \eqref{eq:SEEs} with the density operator yields quasiprobabilities which include negative contributions for entanglement \cite{SV09a}.

	Further, the Hamiltonian can decomposed into local parts for $A$ and $B$ and an interaction contribution,
	\begin{align}
		\hat H=\hat H_A\otimes\hat 1_B+\hat 1_A\otimes\hat H_B+\hat H^\mathrm{(int)}.
	\end{align}
	For example, we can show that for a vanishing interaction, $\hat H^\mathrm{(int)}=0$, the SSEs \eqref{eq:SSEa} and \eqref{eq:SSEb} decouple.
	Namely, we get $i\hbar|\dot a\rangle=\hat H_A|a\rangle$ and $i\hbar|\dot b\rangle=\hat H_B|b\rangle$, being independent SEs [cf. Eq. \eqref{eq:SE}] for each subsystem.
	Clearly, this behavior has to be required from any separable evolution, and in fact, it is a direct consequence of the SSEs.

	Finally, we can introduce a local interaction picture to eliminate the influence of the local evolution.
	This means, we can write $|a(t),b(t)\rangle=\hat U(t)|\tilde a(t),\tilde b(t)\rangle$, with the local unitary $\hat U(t)=e^{-i\hat H_At/\hbar}\otimes e^{-i\hat H_Bt/\hbar}$.
	This leads to SSEs for $|\tilde a\rangle$ and $|\tilde b\rangle$ which depend on the effective Hamiltonian $\hat H^\mathrm{(eff)}(t)=\hat U^\dag(t)\hat H^\mathrm{(int)}\hat U(t)$ only.
	This allows us to focus on the interaction part of the Hamiltonian, since the local parts cannot lead to entanglement.

\begin{figure*}
	\includegraphics[width=\textwidth]{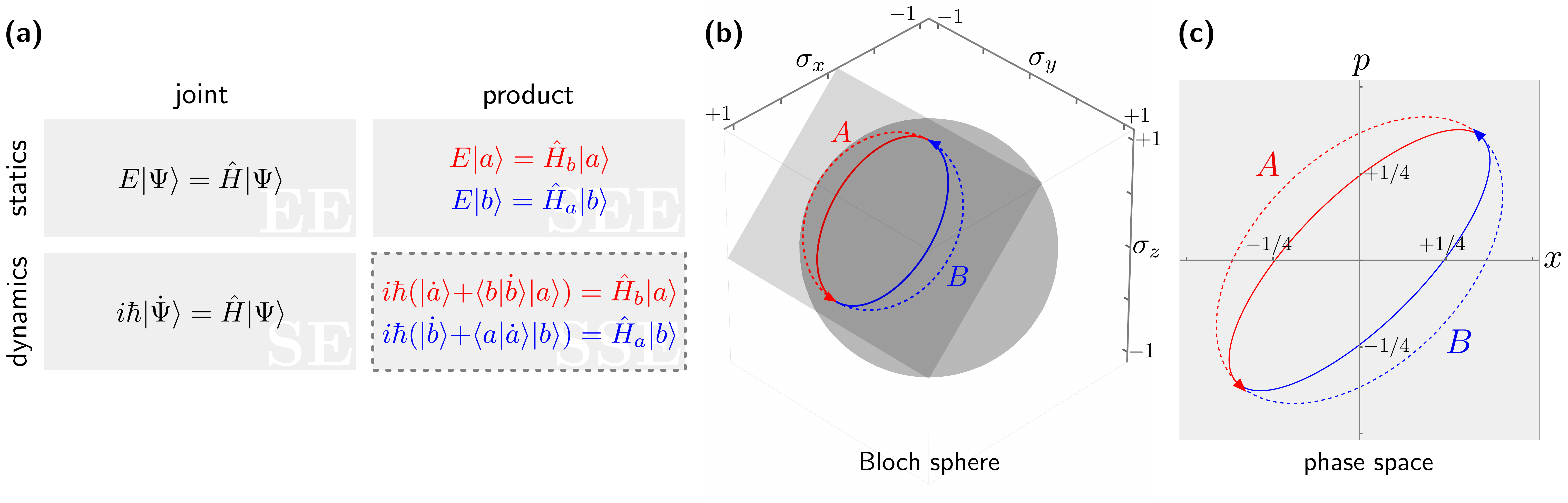}
	\caption{(Color online)
		(a) The relation between the different equations.
			The time-independent equation for the joint system is the EE [Eq. \eqref{eq:EE}].
			For product states, $|a\rangle\otimes|b\rangle$, the static equations are the SEEs [Eq. \eqref{eq:SEEs}].
			The joint evolution of the system is described via the SE [Eq. \eqref{eq:SE}].
			Here, we derived the ``missing link''---the SSEs [Eqs. \eqref{eq:SSEa} and \eqref{eq:SSEb}]---which describe the propagation of product states.
		(b) Trajectories on the Bloch sphere.
			The solutions of the SSE (dashed) are compared to the solutions of the SE (solid) for one half period.
			All curves are on the depicted plane.
			The initial conditions are $|a_0\rangle=|{\uparrow}\rangle$ and $|b_0\rangle=(|{\uparrow}\rangle+|{\downarrow}\rangle)/\sqrt{2}$.
		(c) Trajectories in phase space for one half period.
			The solutions of the SSE and SE are shown as dashed and solid curves, respectively.
			The initial states are the coherent states $|e^{i\pi/4}/2\rangle$ and $|-e^{i\pi/4}/2\rangle$ for the subsystems $A$ and $B$, respectively.
	}\label{fig}
\end{figure*}

	Thus, we proved a number of fundamental features of the SSEs.
	For instance, the static solutions of the SSEs yield the SEE; see Fig. \ref{fig}(a) for a comparison with the SE and the EE.
	Let us apply our approach in the following.

\paragraph*{Application: Exchange interaction.---}\hspace*{-3ex}
	We can focus on the interaction between the subsystems, $\hat H_A=0=\hat H_B$.
	We study a Hamiltonian $\hat H$ which describes an exchange interaction,
	\begin{align}\label{eq:ExchangeHamiltonian}
		\hat H=\hbar\kappa\hat V,
	\end{align}
	where $\hat V$ is the swap operator, $\hat V|a,b\rangle=|b,a\rangle$ for all $|a,b\rangle$, and $\kappa$ is a coupling constant.
	To compare the actual dynamics with the evolution in a separable scenario, we use the initial conditions $|\psi(0)\rangle=|a_0,b_0\rangle$.
	Then the solution of the SE yields
	\begin{align}\label{eq:SolutionSE}
		|\psi(t)\rangle=\cos(\kappa t)|a_0,b_0\rangle-i\sin(\kappa t)|b_0,a_0\rangle.
	\end{align}
	To exclude the trivial case, we assume that the initial states are not parallel, $|q|\neq1$, with
	\begin{align}
		q=\langle a_0|b_0\rangle
	\end{align}
	being the transition amplitude between the initial states.

	We can solve the SSE for the Hamiltonian \eqref{eq:ExchangeHamiltonian} exactly, cf. \cite{supplement}.
	The solutions for the given initial conditions read
	\begin{subequations}
	\begin{align}\label{eq:SolutionSSEa}
		|a(t)\rangle & =\cos(|q|\kappa t)|a_0\rangle-i\frac{q^\ast}{|q|}\sin(|q|\kappa t)|b_0\rangle
	\intertext{and}\label{eq:SolutionSSEb}
		|b(t)\rangle & =\cos(|q|\kappa t)|b_0\rangle-i\frac{q}{|q|}\sin(|q|\kappa t)|a_0\rangle.
	\end{align}
	\end{subequations}
	Let us now compare the solution of the SE [Eq. \eqref{eq:SolutionSE}] with the solutions of the SSEs [Eqs. \eqref{eq:SolutionSSEa} and \eqref{eq:SolutionSSEb}].

	The state $|\psi(t)\rangle$ in Eq. \eqref{eq:SolutionSE} oscillates between the initial state $|a_0,b_0\rangle$ and the swapped state $|b_0,a_0\rangle$, and it is separable solely for the times $t$ which are an integer multiple of $\pi/|2\kappa|$ \cite{supplement}.
	The separable solution $|a(t),|b(t)\rangle$ also oscillates between $|a_0,b_0\rangle$ and $|b_0,a_0\rangle$.
	However, the periods of this oscillation differ from the entangled case.
	The period of the state \eqref{eq:SolutionSE} is $2\pi/|\kappa|$, whereas the period of the separable solutions \eqref{eq:SolutionSSEa} and \eqref{eq:SolutionSSEb} is increased to $2\pi/|\kappa q|$.
	This becomes most significant in the limit $|q|\to 0$, which gives the time-independent solution $|a(t),b(t)\rangle=|a_0,b_0\rangle$.
	Note that $|a_0\rangle\perp |b_0\rangle$ is a solution of the SEEs \eqref{eq:SEEs}, describing the static scenario for the given Hamiltonian \cite{supplement}.
	In the classical picture of Lagrangian mechanics, we can understand the increased time scale through an additional virtual work to be done, which is due to constraining the solutions of the SSE to separable states.
	Clearly, this does not occur for the unconstrained evolution in terms of the SE.
	Consequently, the entangled evolution and the separable evolution under study are described on different time scales.

	As a first example, we assume that each subsystem describes a spin-$1/2$ particle.
	Hence, the system realizes a discrete-variable qubit \cite{NC00}, where the spin-up state $|{\uparrow}\rangle$ (spin-down state $|{\downarrow}\rangle$) represents the truth value ``true'' (``false'').
	Using the Pauli matrices \cite{comment2}, we can write the spin operators for the particles as
	\begin{align}
		\vec{\hat S}_A=\frac{\hbar}{2}\begin{pmatrix}
			\hat\sigma_x\otimes\hat\sigma_0\\
			\hat\sigma_y\otimes\hat\sigma_0\\
			\hat\sigma_z\otimes\hat\sigma_0
		\end{pmatrix}
		\text{ and }
		\vec{\hat S}_B=\frac{\hbar}{2}\begin{pmatrix}
			\hat\sigma_0\otimes\hat\sigma_x\\
			\hat\sigma_0\otimes\hat\sigma_y\\
			\hat\sigma_0\otimes\hat\sigma_z
		\end{pmatrix}.
	\end{align}
	The scalar product of these vectors allows us to bring the exchange-interaction Hamiltonian \eqref{eq:ExchangeHamiltonian} into the form
	\begin{align}
		\hat H=\frac{\hbar\kappa}{2}\hat\sigma_0\otimes\hat\sigma_0+\frac{2\kappa}{\hbar}\vec{\hat S}_A\cdot\vec{\hat S}_B,
	\end{align}
	which yields a spin-spin coupling, with many applications in chemical physics, nuclear physics, and beyond; see, e.g., Ref. \cite{SM99}.
	One representation of qubits is given in terms of the Bloch sphere \cite{NC00}.
	In our case, we assign the tuple
	$(\sigma_x,\sigma_y,\sigma_z)=(\langle \hat \sigma_x\otimes\hat\sigma_0\rangle,\langle \hat \sigma_y\otimes\hat\sigma_0\rangle,\langle \hat \sigma_z\otimes\hat\sigma_0\rangle)$ to the particle $A$ and
	$(\sigma_x,\sigma_y,\sigma_z)=(\langle \hat\sigma_0\otimes\hat \sigma_x\rangle,\langle \hat\sigma_0\otimes\hat \sigma_y\rangle,\langle \hat\sigma_0\otimes\hat \sigma_z\rangle)$ to $B$.
	Note that the operators present local observables, having a form $\hat M_A\otimes\hat M_B$, which typically cannot be applied to verify entanglement.

	In Fig. \ref{fig}(b), we compare the separable trajectories (dashed) with the inseparable trajectories (solid) for one cycle which converts $|a_0,b_0\rangle$ into $|b_0,a_0\rangle$.
	The inseparable trajectory is squeezed compared to the separable one.
	This means that even without a reference time, entanglement of the evolved states is verified by its path on the Bloch sphere which is incompatible with the separable propagation.
	We should also keep in mind that in both cases we start with the same initial conditions.

	Our second example uses the representation in the continuous-variable phase space \cite{A13}, based on generalized position $\hat x$, or quadrature, and its conjugate momentum $\hat p$.
	In quantum optics, a single-mode radiation field may be described via the bosonic annihilation and creation operators $\hat c$ and $\hat c^\dag$, respectively, leading to
	\begin{align}
		\hat x=\frac{\hat c+\hat c^\dag}{2}
		\quad\text{and}\quad
		\hat p=\frac{\hat c-\hat c^\dag}{2i}.
	\end{align}
	Hence, we obtain the phase-space coordinates $(x,p)=(\mathrm{Re}\langle\hat a\rangle,\mathrm{Im}\langle\hat a\rangle)$ for mode $A$, using $\hat a=\hat c\otimes\hat 1$ and $\hat 1_A=\hat 1=\hat 1_B$, and $(x,p)=(\mathrm{Re}\langle\hat b\rangle,\mathrm{Im}\langle\hat b\rangle)$ for mode $B$, with $\hat b=\hat 1\otimes\hat c$.
	To determine the meaning of the exchange interaction, let us recall that $\langle\alpha,\beta|\hat V|\alpha,\beta\rangle=e^{-|\alpha-\beta|^2}=\langle \alpha,\beta|{:} \exp(-[\hat a-\hat b]^\dag[\hat a-\hat b]) {:}|\alpha,\beta\rangle$, where $|\alpha,\beta\rangle$ is a two-mode coherent state and ${:}\cdots{:}$ denotes the normal ordering prescription \cite{VW06}.
	Thus, to get an idea of the operation of the Hamiltonian \eqref{eq:ExchangeHamiltonian}, we can approximate $\hat H$ in a first-order Taylor expansion as \cite{comment3}
	\begin{align}
		\hat H\approx
		\hbar\kappa(\hat 1\otimes\hat 1-\hat a^\dag\hat a-\hat b^\dag\hat b)
		+\hbar\kappa(\hat a^\dag\hat b+\hat b^\dag\hat a).
	\end{align}
	The first terms are local contributions only.
	The second part, however, describes a beam splitter, interfering the optical modes $A$ and $B$, which is vital for realizing quantum-optical experiments, such as photon antibunching \cite{KDM77}, Hong-Ou-Mandel interferences \cite{HOM87}, or more general multiphoton correlations \cite{AKJMSHRWJ17,MJMTBKW17}.

	The trajectories in phase space are shown in Fig. \ref{fig}(c) for $A$ and $B$.
	The curves depict one half period for initially coherent states.
	Similarly to the previous case, the inseparable path is characterized by describing a narrower oval compared to the separable one.
	Again, the relation to classical Lagrangian mechanics identifies the additional virtual work as the reason for the more extend trajectory for separable states---i.e., confined degrees of freedom yield additional virtual work, which corresponds to a larger amplitude in phase space.

\paragraph*{Generalization and outlook.---}\hspace*{-3ex}
	Beyond bipartite systems, entanglement in multipartite systems has a much richer structure and, therefore, a higher complexity \cite{HHHH09}.
	Still, we can generalize the SSEs to the multipartite scenario \cite{supplement}.
	For example, the von Neumann form [cf. Eq. \eqref{eq:vonNeumann}] for a $N$-partite product state $|a_1,\ldots, a_N\rangle$ is given by
	\begin{align}\label{eq:Multipartite}
		i\hbar\frac{d}{dt}\left(|a_n\rangle\langle a_n|\right)=\big[\hat H_{a_1,\ldots,a_{n-1},a_{n+1},\ldots,a_N},|a_n\rangle\langle a_n|\big],
	\end{align}
	for $n=1,\ldots,N$ and where $\hat H_{a_1,\ldots,a_{n-1},a_{n+1},\ldots,a_N}$ is the multipartite generalization of $\hat H_a$ \cite{SV13,supplement}.
	Interestingly, for $N=1$, we retrieve the SE.
	Equation \eqref{eq:Multipartite} allows us to study entangling properties of multipartite quantum dynamics.

	Beyond the fundamental introduction of separable trajectories for pure states, another step is the treatment of mixed separable states \cite{W89}, which can be achieved via mixtures of pure-state trajectories distributed according to the probability distribution which describes the initial state \cite{comment4}.
	In this context, it would be also interesting to study open quantum systems.
	For instance, a separable form of the Lindblad master equations would allow us to infer inseparable trajectories including attenuations.
	This extension of our theory requires further studies, but it might lead to a deeper insight into inseparable quantum trajectories for a broader class of systems.

\paragraph*{Conclusions.---}\hspace*{-3ex}
	We derived equations of motion which render it possible to discern inseparable trajectories from separable ones.
	Using the least action principle, we develop the method of separability Schr\"odinger equations, resembling the original Schr\"odinger equation constrained to separable states.
	We characterized our coupled set of nonlinear equations.
	We found an interesting symmetry between the time-independent and time-dependent Sch\"odinger equation and their corresponding counterparts for separability.
	We also formulated a multipartite generalization of our technique.

	One interpretation of our results is that the action serves as a witness for time-dependent entanglement.
	The action for the separable trajectory has be larger than the action for the inseparable one, which is the global (unrestricted) minimum.
	The difference between those quantities can also serve as a measure to quantitatively assess the generated entanglement.

	Moreover, our method enables us to compare the separable trajectory to the inseparable one, which was impossible before.
	Whenever the inseparable propagations deviates from the evolution of a classically correlated scenario, an entangling dynamics is uncovered.
	This also describes the experimental implementation of our approach.
	Namely, the separable quantum trajectories predicted by our theory can be directly compared to the measured behavior of an interacting system to certify temporal quantum correlations.

	Our method describes the joint, but separable, evolution of quantum systems without disregarding the interaction.
	As an application, we solved our equations of motion for an exchange-interaction Hamiltonian.
	This enabled us to compare the separable dynamics, due to the separability Schr\"odinger equation, with the actual evolution in terms of the Schr\"odinger equation.
	For example, we found that the propagation in time is slower for the separable case.
	Moreover, we studied the quantum trajectories on the discrete-variable Bloch sphere and in the continuous-variable phase space.

	Therefore, we developed a universally applicable concept to identify time-dependent entanglement by considering the whole trajectory---instead of considering a single point in time or restricting ourselves to static scenarios.
	The technique introduced here allows us to not only to predict, but also to certify temporal forms of inseparability.
	We believe that this method is a step towards novel applications and a deeper fundamental understanding of time-dependent quantum phenomena.

\paragraph*{Acknowledgments.---}\hspace*{-3ex}
	The project leading to this application has received funding from the European Union's Horizon 2020 research and innovation program under the grant agreement No. 665148 (QCUMbER).


\section*{Appendices}
	We briefly review the separability eigenvalue equations in Sec. \ref{App:sec:SEEs}.
	We prove properties of the separability Schr\"odinger equations (SSEs) in Sec. \ref{App:sec:Properties}.
	The exact solution of the considered example is given in Sec. \ref{App:sec:Example}.
	Finally, the multipartite SSEs are derived in Sec. \ref{App:sec:Multipartite}, and some comments on time-dependent and global phases are provided in Sec. \ref{App:sec:Phase}.

\appendix

\section{Separability eigenvalue equations}\label{App:sec:SEEs}
	The bipartite separability eigenvalue equations for an operator $\hat L$ read \cite{SV09}
	\begin{align}
		\hat L_b|a\rangle=g|a\rangle
		\quad\text{and}\quad
		\hat L_a|b\rangle=g|b\rangle,
	\end{align}
	where $\hat L_{a}=\mathrm{tr}_A[\hat L(|a\rangle\langle a|\otimes\hat 1_B)]$ and $\hat L_{b}=\mathrm{tr}_B[\hat L(\hat 1_A\otimes|b\rangle\langle b|)]$.
	The normalized vectors $|a\rangle$ and $|b\rangle$ form the separability eigenvector $|a,b\rangle=|a\rangle\otimes|b\rangle$.
	The number $g$ is the separability eigenvalue.
	A multipartite generalization of these equations were also formulated \cite{SV13}.
	For comparison, the standard eigenvalue equation reads $\hat L|\psi\rangle=g|\psi\rangle$.

	The operator $\hat L$ is completely determined if all expectation values of the form $\langle x,y|\hat L|x,y\rangle$ are known for a complex, tensor-product Hilbert space.
	We have
	\begin{align}\label{App:eq:PartReducedOpsRel}
		\langle x,y|\hat L|x,y\rangle=\langle x|\hat L_y|x\rangle=\langle y|\hat L_x|y\rangle.
	\end{align}
	It well known that $\langle x,y|\hat L|x,y\rangle=\langle x,y|\hat L|x,y\rangle^\ast$ for all $|x,y\rangle$ in a complex, tensor-product Hilbert space is identical to the statement that $\hat L$ is Hermitian.
	Therefore, the definition of $\hat L_x$ and $\hat L_y$ implies for any operator $\hat L=\hat L^\dag$ that
	\begin{align}
		\hat L_x=(\hat L_x)^\dag
		\quad\text{and}\quad
		\hat L_y=(\hat L_y)^\dag.
	\end{align}

\section{Properties of the SSE}\label{App:sec:Properties}
	In this section, we study properties of the dynamics to be inferred from the SSEs,
	\begin{subequations}\label{App:eq:SSEs}
	\begin{align}
		\label{App:eq:SSEa}
		i\hbar\left(\langle b|b\rangle|\dot a\rangle+\langle b|\dot b\rangle |a\rangle \right)=&\hat H_b|a\rangle
		\intertext{and}
		\label{App:eq:SSEb}
		i\hbar\left(\langle a|a\rangle|\dot b\rangle+\langle a|\dot a\rangle |b\rangle \right)=&\hat H_a|b\rangle.
	\end{align}
	\end{subequations}
	for
	\begin{align}
		|a\rangle=|a(t)\rangle
		\quad\text{and}\quad
		|b\rangle=|b(t)\rangle
	\end{align}
	Furthermore, we define $E=\langle a,b|\hat H|a,b\rangle$, and the initial states are labeled as $|a(0)\rangle=|a_0\rangle$ and $|b(0)\rangle=|b_0\rangle$, with $\langle a_0|a_0\rangle=1=\langle b_0|b_0\rangle$.
	The Hamiltonian is decomposed in the form
	\begin{align}\label{App:eq:DecomposedHamiltonian}
		\hat H=\hat H_{A}\otimes\hat 1_B+\hat 1_{A}\otimes\hat H_B+\hat H^\mathrm{(int)}.
	\end{align}

\subsection{Conservation of normalization}
	Multiplying Eq. \eqref{App:eq:SSEa} with $\langle a|$ or Eq. \eqref{App:eq:SSEb} with $\langle b|$ results in
	\begin{align}
		i\hbar\left(\langle b|\dot b\rangle \langle a|a\rangle+\langle b|b\rangle\langle a|\dot a\rangle\right)=E
	\end{align}
	Because $\hat H=\hat H^\dag$ implies $E=E^\ast$ for all $|a,b\rangle$, we find
	\begin{align}
		&\frac{d\langle a,b|a,b\rangle}{dt}
		=-\frac{E^\ast}{i\hbar}+\frac{E}{i\hbar}=0.
	\end{align}
	Since the initial state has a total probability of one, $\langle a(0),b(0)|a(0),b(0)\rangle=1$, we get
	\begin{align}
		\langle a(t),b(t)|a(t),b(t)\rangle=1
	\end{align}
	for all $t$.
	Hence, we can write for all times
	\begin{align}
		|a,b\rangle=\frac{|a,b\rangle}{\sqrt{\langle a,b|a,b\rangle}}=\frac{|a\rangle}{\sqrt{\langle a|a\rangle}}\otimes\frac{|b\rangle}{\sqrt{\langle b|b\rangle}}.
	\end{align}
	Therefore, we can state without a loss of generality that the states of the subsystems have a constant normalization.
	That is, $\langle a|a\rangle=1=\langle b|b\rangle$ holds true for all times.

\subsection{\lowercase{von} Neumann form}
	To derive the von Neumann form of the SSEs, we can compute from the original SSEs for subsystem $A$
	\begin{align}
		\nonumber
		\frac{d}{dt}(|a\rangle\langle a|)
		=&\left(\frac{1}{i\hbar}\hat H_b|a\rangle-\langle b|\dot b\rangle |a\rangle\right)\langle a|
		\\\nonumber
		&+|a\rangle\left(-\frac{1}{i\hbar}\langle a|\hat H_b-\langle \dot b|b\rangle\langle a|\right)
		\\
		=&\frac{1}{i\hbar}[\hat H_b,|a\rangle\langle a|]-\underbrace{\left(\langle b|\dot b\rangle+\langle \dot b|b\rangle\right)}_{\displaystyle =\frac{d\langle b|b\rangle}{dt}=0}|a\rangle\langle a|,
	\end{align}
	and similarly for subsystem $B$.
	Thus, we have
	\begin{subequations}\label{App:eq:vonNeumann}
	\begin{align}
		i\hbar\frac{d}{dt}(|a\rangle\langle a|)&=[\hat H_b,|a\rangle\langle a|],
		\\
		i\hbar\frac{d}{dt}(|b\rangle\langle b|)&=[\hat H_a,|b\rangle\langle b|].
	\end{align}
	\end{subequations}
	Note that these von Neumann-type equations are equivalent to the SSEs when ignoring global phases.
	A similar relation holds for the Schr\"odinger equation, where $i\hbar\,d|\psi\rangle/dt=\hat H|\psi\rangle$ is also equivalent to $i\hbar\, d(|\psi\rangle\langle \psi|)/dt=[\hat H,|\psi\rangle\langle\psi|]$ up to a global phase.

\subsection{Conservation of energy}
	Let us consider the expectation value of a time-independent observable $\hat L$.
	From the von Neumann form of the SSEs, we get
	\begin{align}
		i\hbar\frac{d}{dt}\langle a,b|\hat L|a,b\rangle
		=\langle a|[\hat L_b,\hat H_b]|a\rangle
		+\langle b|[\hat L_a,\hat H_a]|b\rangle.
	\end{align}
	For instance, this time derivative becomes zero if $[\hat L_b,\hat H_b]=0$ and $[\hat L_a,\hat H_a]=0$.
	In particular, this is true for $\hat L=\hat H$, resulting in
	\begin{align}
		\frac{dE}{dt}=0.
	\end{align}

\subsection{Time-independent solutions}
	In the stationary case with vanishing time derivatives, the von Neumann-type Eq. \eqref{App:eq:vonNeumann} take the forms $0=[\hat H_b,|a\rangle\langle a|]$ and $0=[\hat H_a,|b\rangle\langle b|]$.
	This is equivalent to $|a\rangle$ and $|b\rangle$ being eigenvectors of $\hat H_b$ and $\hat H_a$, respectively.
	This means that the stationary case yields separability eigenvectors,
	\begin{align}
		\hat H_b|a\rangle=E|a\rangle
		\quad\text{and}\quad
		\hat H_a|b\rangle=E|b\rangle,
	\end{align}
	where $E=\langle a,b|\hat H|a,b\rangle$ is applied to determine that the eigenvalue is the energy.

\subsection{Noninteracting systems}
	Here, we assume that the subsystems do not interact, $\hat H^\mathrm{(int)}=0$ [see Eq. \eqref{App:eq:DecomposedHamiltonian}].
	Then the partially reduced Hamiltonian are simply $\hat H_a=\hat H_B+\langle a|\hat H_A|a\rangle\hat 1_B$ and $\hat H_b=\hat H_A+\langle b|\hat H_B|b\rangle\hat 1_A$.
	Hence, the equations of motion decouple,
	\begin{align}
	\begin{aligned}
		i\hbar\frac{d}{dt}(|a\rangle\langle a|)=[\hat H_A,|a\rangle\langle a|]
		\\\text{and}\quad
		i\hbar\frac{d}{dt}(|b\rangle\langle b|)=[\hat H_B,|b\rangle\langle b|],
	\end{aligned}
	\end{align}
	which is equivalent (ignoring a global phase) to the local Schr\"odinger equations
	\begin{align}
		i\hbar \frac{d}{dt}|a\rangle=\hat H_A|a\rangle
		\quad\text{and}\quad
		i\hbar \frac{d}{dt}|b\rangle=\hat H_B|b\rangle.
	\end{align}

\subsection{Interaction picture}
	For studying the case of a nonvanishing interaction, $\hat H^\mathrm{(int)}\neq0$, we can further introduce
	\begin{align}
		|a\rangle=e^{-i\hat H_A t/\hbar}|x\rangle
		\quad\text{and}\quad
		|b\rangle=e^{-i\hat H_B t/\hbar}|y\rangle.
	\end{align}
	From this we get the time derivatives, such as
	\begin{align}
	\begin{aligned}
		i\hbar |\dot a\rangle=&e^{-i\hat H_A t/\hbar}\left(\hat H_A|x\rangle+|\dot x\rangle\right)
		\\\text{and}\quad
		i\hbar \langle b|\dot b\rangle=&\langle y|\hat H_B|y\rangle+\langle y|\dot y\rangle,
	\end{aligned}
	\end{align}
	and the action of the partially reduced Hamiltonian,
	\begin{align}
		\hat H_b|a\rangle=&e^{-i\hat H_A t/\hbar}\left(\hat H_A|x\rangle+\langle y|\hat H_B|y\rangle|x\rangle+\hat H^\mathrm{(eff)}_y|x\rangle\right),
	\end{align}
	using the locally transformed interaction part of the considered Hamiltonian,
	\begin{align}\nonumber
		&\hat H^\mathrm{(eff)}
		\\\label{App:eq:effH}
		=&\!\left(e^{i\hat H_A t/\hbar}\otimes e^{i\hat H_B t/\hbar}\right)\hat H^\mathrm{(int)}\left(e^{-i\hat H_A t/\hbar}\otimes e^{-i\hat H_B t/\hbar}\right)\!.
	\end{align}
	\begin{subequations}
	Thus, we get for the SSE for subsystem $A$
	\begin{align}
		i\hbar\left(|\dot x\rangle+\langle y|\dot y\rangle|x\rangle\right)=\hat H^\mathrm{(eff)}_y|x\rangle,
	\end{align}
	applying the normalization $\langle a|a\rangle=\langle x|x\rangle=1=\langle y|y\rangle=\langle b|b\rangle$.
	Similarly, we obtain for subsystem $B$
	\begin{align}
		i\hbar\left(|\dot y\rangle+\langle x|\dot x\rangle|y\rangle\right)=\hat H^\mathrm{(eff)}_x|y\rangle.
	\end{align}
	\end{subequations}
	Hence, we obtain a new set of SEEs for $|x,y\rangle$ and the effective Hamiltonian \eqref{App:eq:effH}.

\section{Exchange interaction}\label{App:sec:Example}
	In this section, we provide the exact results and supplemental discussion for the application studied.

\subsection{The standard approach}
	We consider the the Hamiltonian
	\begin{align}
		\hat H=\hbar\kappa\hat V,
	\end{align}
	where $\kappa$ is a real-valued constant and $\hat V$ is the swap operator, $\hat V|x,y\rangle=|y,x\rangle$ for all $|x\rangle$ and $|y\rangle$.
	The eigenvalues of $\hat H$ are $\pm\hbar\kappa$, which can be seen from $\hat V(|x,y\rangle\pm|y,x\rangle)=\pm(|x,y\rangle\pm|y,x\rangle)$.
	For convenience, we use a rescaled and unit-free time,
	\begin{align}
		\tau=\kappa t.
	\end{align}
	The straightforward solution of the standard Schr\"odinger equation for the initial state $|\psi(0)\rangle=|a_0,b_0\rangle$ yields the following propagated state:
	\begin{align}
		|\psi(\tau)\rangle=\cos(\tau)|a_0,b_0\rangle-i\sin(\tau)|b_0,a_0\rangle.
	\end{align}
	From the Schmidt decomposition (see Ref. \cite{NC00}), we get time-dependent Schmidt coefficients,
	\begin{align}
		\lambda_\pm=\sqrt{\frac{1\pm\sqrt{1-\sin^2(2\tau)(1-|q|^2)^2}}{2}},
	\end{align}
	where we use the transition amplitude
	\begin{align}\label{App:eq:Overlap}
		q=\langle a_0|b_0\rangle.
	\end{align}
	We have $\lambda_-=0$---resembling the separable case---for parallel initial states ($|q|=1$) or for any time $\tau$ which is an integer-multiple of $\pi/2$.

\subsection{Solving the SSEs}
	Now we analytically solve the nonlinear SSEs.
	From $\langle y|\hat V_x|y\rangle=\langle y|x\rangle\langle x|y\rangle$ for all $|y\rangle$, we can conclude that the reduced operator $\hat V_x$ has the form
	\begin{align}\label{eq:ReducedSwapOperator}
		\hat V_x=|x\rangle\langle x|.
	\end{align}
	Thus, we have that $\hat V_x|y\rangle=0|y\rangle$ for $|y\rangle$ perpendicular to $|x\rangle$ and  $\hat V_x|y\rangle=1|y\rangle$ for $|y\rangle$ parallel to $|x\rangle$.
	Hence, the separability eigenvalues of $\hat H$ are $0$ and $\hbar\kappa$.
	Note, it holds that $\langle x,y|\hat V|x,y\rangle=|\langle x|y\rangle|^2\geq0$.

	Therefore, we can write for the von Neumann form
	\begin{align}\label{eq:ExampleVonNeumann}
		i\frac{d}{d\tau}(|a\rangle\langle a|)=[|b\rangle\langle b|,|a\rangle\langle a|]=-i\frac{d}{d\tau}(|b\rangle\langle b|).
	\end{align}
	This implies that the operator $\hat C=|b\rangle\langle b|+|a\rangle\langle a|$ is time-independent, $id\hat C/d\tau=0$.
	To satisfy the initial conditions, $|a_0,b_0\rangle$ for $\tau=0$, $\hat C$ has to obey
	\begin{align}
		\hat C=|b_0\rangle\langle b_0|+|a_0\rangle\langle a_0|.
	\end{align}
	When replacing $|b\rangle\langle b|=\hat C-|a\rangle\langle a|$ in the equation of motion for $|a\rangle\langle a|$, we can obtain the solutions from the standard theory.
	That is, the von Neumann-type equation $id(|c\rangle\langle c|)/d\tau=[\hat C,|c\rangle\langle c|]$,
	with
	\begin{align}
		|c(\tau)\rangle=\gamma(\tau)|a_0\rangle+\delta(\tau)|b_0\rangle,
	\end{align}
	is equivalent to the Schr\"odinger-type equation
	\begin{align}
		i\frac{d}{d\tau}|c\rangle=\hat C|c\rangle.
	\end{align}
	Here it is also worth recalling that $\hat C=|a_0\rangle\langle a_0|+|b_0\rangle\langle b_0|$.

	The latter Schr\"odinger-type equation is solved for the time-dependent parameters
	\begin{align}
		\!\begin{pmatrix}\gamma(\tau)\\\delta(\tau)\end{pmatrix}\!
		=\!\begin{pmatrix}
			\cos(|q|\tau) & -i\frac{q^\ast}{|q|}\sin(|q|\tau) \\
			-i\frac{q^\ast}{|q|}\sin(|q|\tau) & \cos(|q|\tau)
		\end{pmatrix}\!\!
		\begin{pmatrix}\gamma(0)\\\delta(0)\end{pmatrix}\!,
	\end{align}
	using $q$ as defined in Eq. \eqref{App:eq:Overlap} and while ignoring a global phase $e^{-i\tau}$.
	The initial conditions $\gamma(0)=1$ and $\delta(0)=0$ [or $\gamma(0)=0$ and $\delta(0)=1$] correspond to the states $|c(0)\rangle=|a_0\rangle$ [or $|c(0)\rangle=|b_0\rangle$].
	Therefore, we finally get the separable trajectories of the exchange interaction as
	\begin{subequations}
	\begin{align}
		|a(\tau)\rangle=&\cos(|q|\tau)|a_0\rangle-i\frac{q^\ast}{|q|}\sin(|q|\tau)|b_0\rangle,
		\\
		|b(\tau)\rangle=&\cos(|q|\tau)|b_0\rangle-i\frac{q}{|q|}\sin(|q|\tau)|a_0\rangle.
	\end{align}
	\end{subequations}

\section{Multipartite separable evolution}\label{App:sec:Multipartite}
	In the $N$-partite case, pure separable states take the form
	\begin{align}
		|\psi^\mathrm{(sep)}\rangle=|a_1,\ldots,a_N\rangle=\bigotimes_{n=1}^N |a_n\rangle.
	\end{align}
	For each subsystem, $l=1,\ldots,N$, the Euler-Lagrange equation reads
	\begin{align}\label{App:eq:EulerLagrange}
		0=&\frac{d}{dt}\frac{\partial L}{\partial \langle \dot a_l|}-\frac{\partial L}{\partial \langle a_l|}.
	\end{align}
	The time derivative of $|\psi^\mathrm{(sep)}\rangle$ can be obtained with the product rule,
	\begin{align}
		\frac{d}{dt}|\psi^\mathrm{(sep)}\rangle
		=
		\sum_{m=1}^N
		\left[\bigotimes_{n=1}^{m-1}|a_n\rangle\right]
		\otimes|\dot a_m\rangle\otimes
		\left[\bigotimes_{n=m+1}^{N}|a_n\rangle\right],
	\end{align}
	where we identify $|\dot a_m\rangle=d|a_m\rangle/dt$.
	Hence, the Lagrangian for this $N$-partite product state reads
	\begin{align}
		\nonumber
		L=&
		\frac{i\hbar}{2}\langle \psi^\mathrm{(sep)}|\left[\frac{d}{dt}|\psi^\mathrm{(sep)}\rangle\right]
		-\frac{i\hbar}{2}\left[\frac{d}{dt}\langle\psi^\mathrm{(sep)}|\right]|\psi^\mathrm{(sep)}\rangle
		\\\nonumber
		&-\langle \psi^\mathrm{(sep)}|\hat H|\psi^\mathrm{(sep)}\rangle
		\\\nonumber
		=&\frac{i\hbar}{2}\sum_{m}\left[\prod_{n\neq m}\langle a_n|a_n\rangle\right]
		\left[
			\langle a_m|\dot a_m\rangle
			-\langle \dot a_m|a_m\rangle
		\right]
		\\&-\langle a_1,\ldots,a_N|\hat H|a_1,\ldots,a_N\rangle.
	\end{align}

\begin{widetext}
	We find
	\begin{align}
		\frac{\partial L}{\partial \langle \dot a_l|}=
		&-\frac{i\hbar}{2}\left[\prod_{n\neq l} \langle a_n|a_n\rangle\right]|a_l\rangle,
		\\
		\frac{d}{dt}\frac{\partial L}{\partial \langle \dot a_l|}=
		&-\frac{i\hbar}{2}\left[\prod_{n\neq l} \langle a_n|a_n\rangle\right]|\dot a_l\rangle
		-\frac{i\hbar}{2}\sum_{m\neq l}
		\left[\prod_{n\neq m,l}\langle a_n|a_n\rangle\right]
		\left[\langle \dot a_m|a_m\rangle+\langle a_m|\dot a_m\rangle\right]
		|a_l\rangle,
		\\
		\frac{\partial L}{\partial \langle a_l|}=
		&\frac{i\hbar}{2}\left[\prod_{n\neq l} \langle a_n|a_n\rangle\right]|\dot a_l\rangle
		+\frac{i\hbar}{2}
			\sum_{m\neq l}\left[\prod_{n\neq m,l}\langle a_n|a_n\rangle\right]
			\left[\langle a_m|\dot a_m\rangle-\langle \dot a_m|a_m\rangle\right]
		|a_l\rangle
		-\hat H_{a_1,\ldots,a_{l-1},a_{l+1},\ldots,a_N}|a_l\rangle.
	\end{align}
	where we use the multipartite partially reduced operators $\hat H_{a_1,\ldots,a_{l-1},a_{l+1},\ldots,a_N}$ \cite{SV13},
	\begin{align}
		\hat H_{a_1,\ldots,a_{l-1},a_{l+1},\ldots,a_N}
		=\mathrm{tr}_{1}\cdots\mathrm{tr}_{l-1}\mathrm{tr}_{l+1}\cdots\mathrm{tr}_{N}\bigg[\hat H\bigg(
			|a_1\rangle\langle a_1|\otimes\cdots\otimes|a_{l-1}\rangle\langle a_{l-1}|\otimes\hat 1_l\otimes|a_{l+1}\rangle\langle a_{l+1}|\otimes\cdots\otimes|a_N\rangle\langle a_N|
		\bigg)\bigg].
	\end{align}
	Inserting the derivatives into the $l$th Euler-Lagrange equation [Eq. \eqref{App:eq:EulerLagrange}], we get
	\begin{align}
		0=
		-i\hbar\left[\prod_{n\neq l} \langle a_n|a_n\rangle\right]|\dot a_l\rangle
		-i\hbar
			\sum_{m\neq l}\left[\prod_{n\neq m,l}\langle a_n|a_n\rangle\right]
		\langle a_m|\dot a_m\rangle|a_l\rangle
		+\hat H_{a_1,\ldots,a_{l-1},a_{l+1},\ldots,a_N}|a_l\rangle.
	\end{align}
	\clearpage
\end{widetext}

	\noindent
	From this, we finally obtain ($l=1,\ldots,N$)
	\begin{align}\nonumber
		&i\hbar\left[
			\prod_{n\neq l} \langle a_n|a_n\rangle|\dot a_l\rangle+
			\sum_{m\neq l}\langle a_m|\dot a_m\rangle\prod_{n\neq m,l}\langle a_n|a_n\rangle|a_l\rangle
		\right]
		\\
		=&\hat H_{a_1,\ldots,a_{l-1},a_{l+1},\ldots,a_N}|a_l\rangle,
	\end{align}

	These equations of motion represent the multipartite version of the SSEs.
	The properties, which have been derived for the bipartite case, can be straightforwardly generalized to the multipartite scenario.
	For instance, the normalization of each component of the separable state is constant for all times, $\langle a_n|a_n\rangle=1$ for $n=1,\ldots,N$, and the von Neumann form can be applied.
	Also note that the case $N=1$ yields the standard Schr\"odinger equation, $i\hbar\, d|\psi\rangle/dt=\hat H|\psi\rangle$.

\section{Global time-dependent phases}\label{App:sec:Phase}

	When substituting $|\psi(t)\rangle=e^{Et/(i\hbar)}|\chi(t)\rangle$, the Schr\"odinger equation takes the form
	\begin{align}\label{App:eq:SEDelta}
		i\hbar\frac{d}{dt}|\chi\rangle=\Delta\hat H|\chi\rangle,
	\end{align}
	where $\Delta\hat H=\hat H-E$ and $E=\langle \chi|\hat H|\chi\rangle=\langle \psi|\hat H|\psi\rangle$.
	This also results in
	\begin{align}
		\langle\chi|\dot\chi\rangle=0.
	\end{align}

	This is interesting since the new terms $\langle a|\dot a\rangle$ and $\langle b|\dot b\rangle$ appear in the SSEs \eqref{App:eq:SSEs}.
	From the SSEs and the conservation of normalization, we get $i\hbar(\langle a|\dot a\rangle+\langle b|\dot b\rangle)=E$, using the time-independent energy $E=\langle a,b|\hat H|a,b\rangle$.
	In addition, let us define the function $\varphi(t)=i\hbar(\langle a|\dot a\rangle-\langle b|\dot b\rangle)$.

	For the desired substitution, we consider the states $|x\rangle$ and $|y\rangle$ given by
	\begin{align}\label{App:eq:SubPhase}
		|a\rangle=e^{(Et+\varphi)/(2i\hbar)}|x\rangle
		\quad\text{and}\quad
		|b\rangle=e^{(Et-\varphi)/(2i\hbar)}|y\rangle.
	\end{align}
	The combination of the above definitions yields
	\begin{align}
		\langle x|\dot x\rangle=0
		\quad\text{and}\quad
		\langle y|\dot y\rangle=0.
	\end{align}
	Inserting Eq. \eqref{App:eq:SubPhase} into the SEEs \eqref{App:eq:SSEa} and \eqref{App:eq:SSEb}, we get modified SSEs in the form
	\begin{align}\label{App:eq:SSEDelta}
		i\hbar\frac{d}{dt}|x\rangle=\Delta\hat H_y|x\rangle
		\quad\text{and}\quad
		i\hbar\frac{d}{dt}|y\rangle=\Delta\hat H_x|y\rangle,
	\end{align}
	where $\Delta\hat H_x=\hat H_x-E$ and $\Delta\hat H_y=\hat H_y-E$.

	The form of the SSEs in Eq. \eqref{App:eq:SSEDelta} can be convenient when one does not want to consider global phases.
	In addition, this form of the SSEs shows an even closer resemblance to the Schr\"odinger equation in the form \eqref{App:eq:SEDelta}.
	For the multipartite case, $|a_1,\ldots,a_N\rangle$, we analogously get
	\begin{align}
		i\hbar\frac{d}{dt}|x_l\rangle=\Delta\hat H_{x_1,\ldots,x_{l-1},x_{l+1},\ldots,x_N}|x_l\rangle,
	\end{align}
	for $l=1,\ldots,N$ (see also Sec. \ref{App:sec:Multipartite}).



\begin{thebibliography}{99}
\bibitem{comment1}
		We use a symmetric Lagrangian.
		Another frequently applied form is $L=i\hbar\langle\psi|\dot\psi\rangle-\langle\psi|\hat H|\psi\rangle$ \cite{GR96}.
		The Hamiltonians considered here satisfy $d\hat H/dt=0$ and $\hat H=\hat H^\dag$.
\bibitem{comment2}
		In our notation, the Pauli matrices read
		$\hat\sigma_x=|{\uparrow}\rangle\langle{\downarrow}|+|{\downarrow}\rangle\langle{\uparrow}|$,
		$\hat \sigma_y=i|{\downarrow}\rangle\langle{\uparrow}|-i|{\uparrow}\rangle\langle{\downarrow}|$,
		and $\hat \sigma_z=|{\uparrow}\rangle\langle{\uparrow}|-|{\downarrow}\rangle\langle{\downarrow}|$.
		The identity can be written as $\hat\sigma_0=|{\uparrow}\rangle\langle{\uparrow}|+|{\downarrow}\rangle\langle{\downarrow}|$.
\bibitem{comment3}
		Note, for example, the second-order expansion contains fourth-order contributions of the annihilation and creation operators, which describe nonlinear Kerr-type interactions.
\bibitem{comment4}
		For the initial, mixed state $\hat\sigma(0)=\int dP(a_0,b_0) |a_0,b_0\rangle\langle a_0,b_0|$, we can write the evolved state as $\hat\sigma(t)=\int dP(a_0,b_0)$ $|a(t),b(t)\rangle\langle a(t),b(t)|$, where $|a(t),b(t)\rangle$ satisfies the SSEs [Eqs. \eqref{eq:SSEa} and \eqref{eq:SSEb}] for the initial conditions $|a(0),b(0)\rangle=|a_0,b_0\rangle$.
		When allowing for quasiprobabilities $P$ \cite{SV09a}, the separable dynamics can be studied even for an initially entangled state.
\end{thebibliography}
\end{document}